\documentclass[preprint,12pt,3p]{elsarticle}

\usepackage{multicol}
\usepackage{color}
\usepackage{xspace}
\usepackage{enumerate}
\usepackage{url}
\usepackage[none]{hyphenat}
\usepackage{graphicx}
\usepackage{verbatim}
\usepackage{siunitx}
\usepackage{amssymb}
\usepackage{float}
\usepackage{subcaption}
\usepackage{xcolor}
\usepackage[utf8]{inputenc}
\usepackage{lineno}
\usepackage{textcomp}
\journal{Astroparticle Physics}

\begin{document}

\begin{frontmatter}
\title{Reconstructing air shower parameters with LOFAR using event specific GDAS atmospheres }

\address[1]{Astrophysical Institute, Vrije Universiteit Brussel, Pleinlaan 2, 1050 Brussels, Belgium} 

\address[2]{Department of Astrophysics / IMAPP, Radboud University Nijmegen,  P. O. Box 9010, 6500 GL, Nijmegen, The Netherlands}
\address[3]{NIKHEF, Science Park Amsterdam, 1098 XG Amsterdam, The Netherlands} 
\address[4]{Netherlands Institute of Radio Astronomy (ASTRON), Postbus 2, 7990 AA Dwingeloo, The Netherlands }
\address[5]{KVI-CART, University Groningen, P. O. Box 72, 9700 AB Groningen, The Netherlands}
\address[6]{DESY, Platanenallee 6, 15738 Zeuthen, Germany}
\address[7]{Interuniversity Institute for High-Energy, Vrije Universiteit Brussel,  Pleinlaan 2, 1050 Brussels, Belgium} 
\address[8]{Institut f\"{u}r Kernphysik, Karlsruhe Institute of Technology(KIT), P. O. Box 3640, 76021, Karlsruhe, Germany}
\address[9]{Institut f\"{u}r Physik, Humboldt-Universit\"{a}t zu Berlin, 12489 Berlin, Germany}
\address[10]{Department of Physics, School of Education, Can Tho University Campus II, 3/2 Street, Ninh Kieu District, Can Tho City, Vietnam}
\address[11]{Max-Planck Institute for Radio Astronomy, Bonn, Germany}

\author[1]{P.~Mitra}\corref{cor1}
\ead{pmitra@vub.be}

\author[2,1]{A.~Bonardi}
\author[2]{A.~Corstanje}
\author[1,2]{S.~Buitink}
\author[1]{G.~K Krampah}
\author[2,3,4,11]{H. ~Falcke}
\author[5]{B.~M. ~Hare}
\author[1,2,3]{J. R. ~H\"orandel}
\author[8,1]{T.~Huege}
\author[1]{ K.~Mulrey}
\author[6,9]{A ~Nelles}
\author[1]{H.~Pandya}
\author[1]{J.P.~Rachen}
\author[2]{L.~Rossetto}
\author[5,7]{O.~Scholten}
\author[4]{S.~ter Veen}
\author[5,10]{T.N.G.~Trinh}
\author[1,11]{T.~Winchen}
\cortext[cor1]{Corresponding author}

\begin{abstract}

The limited knowledge of atmospheric parameters like humidity,  
pressure, temperature, and the index of refraction has been one of the important systematic uncertainties in 
reconstructing the depth of the shower maximum
from the radio emission of air showers. Current air shower Monte Carlo simulation codes like 
CORSIKA and the radio plug-in CoREAS use various averaged parameterized atmospheres. 
However, time-dependent and location-specific atmospheric models are needed
for the cosmic ray analysis method used for LOFAR data. There, dedicated simulation
sets are used for each detected cosmic ray, to take into account the actual atmospheric conditions
at the time of the measurement. Using the Global Data Assimilation System (GDAS), a global atmospheric model, we have implemented 
time-dependent, realistic atmospheric profiles in CORSIKA  and CoREAS. \textcolor{black}{We have produced realistic event-specific atmospheres for all air showers 
measured with LOFAR, an event set spanning several years and many different weather conditions. 
A complete re-analysis of our data set shows that for the majority of data, our previous correction factor performed rather well; 
we found only a small systematic shift of 2 g/cm$^2$ in the reconstructed $X_{\rm max}$. However, under extreme weather conditions, for example, very low air pressure, the shift can be up to 15 g/cm$^2$. 
We provide a correction formula to determine the 
shift in $X_{\rm max}$ resulting from a comparison of simulations done using the US-Std atmosphere and the GDAS-based atmosphere.}

\end{abstract}


\end{frontmatter}
\section{Introduction}
In recent years,  the field of radio detection of air 
showers has advanced quite rapidly~\cite{huege2016,Frank}. Estimating the depth 
of the shower maximum, $X_{\rm max}$, with improved accuracy is 
of great interest for the study of the primary particle composition \cite{nature,Apel}. 
The development of the air shower induced by a cosmic ray is governed by the interactions and 
decays of the secondary particles. The secondary electrons and 
positrons in the air shower undergo charge separation as
they travel through the magnetic field of the Earth. This leads to a 
time-varying transverse current, producing radio emission. There is
another small contribution to the radiation from the excess of negative charge accumulated
at the shower front,  known as the \textquoteleft{Askaryan effect}\textquoteright~\cite{Askaryan}.
The emission reaches the ground as a short pulse on the order of 10 to 100 ns 
with a specific lateral intensity distribution, or footprint, that depends 
on $X_{\rm max}$; $X_{\rm max}$  is calculated in terms of total atmospheric matter traversed by the air shower
from the top of the atmosphere to the point where the particle number reaches the maximum.
\textcolor{black}{It is therefore important to know the altitude-dependent air density.}
Another atmospheric parameter that plays a crucial role 
in the  radio emission is the refractive index of air.
If for a given emission region along the shower axis an observer is
located at the corresponding Cherenkov angle, radiation
emitted from all along this region arrives simultaneously.
This results  in a highly compressed signal in time, forming
a ring-like structure on the ground~\cite{anna,ANITA_che}.
The refractive index determines the propagation velocity of the radio signal
at different altitudes and influences the time compression \cite{zharies,krijn}.  
For observers located on the Cherenkov ring, pulses are coherent up to GHz frequencies \cite{Smida}. The angle at which Cherenkov emission is emitted is inversely proportional
to the refractive index. At higher frequencies pulses are more sensitive to the refractive index. In general,  at all frequencies, the variations in the refractive index lead
to changes in the radio intensity footprint \cite{Arthurpaper}. \textcolor{black}{ Both the density and the refractive index of air are dependent on air 
temperature, humidity and pressure. Thus, having a good understanding of these atmospheric variables is crucial. }

\vspace{0.5 cm}
The radio detection technique can be used in combination with established
techniques such as fluorescence detection and surface detection with scintillators and water Cherenkov detectors. Dense antenna arrays like the core
of the LOFAR radio telescope \cite{lofar}\,  provide the 
opportunity to investigate  the radio footprint, i.e.\ the
lateral intensity distribution, in close detail and enable the measurement of
$X_{\rm max}$ up to a precision of $<$ 20 $\mathrm{g/cm ^{2}}$.
The precision is sensitive to the choice of an atmospheric  model included in the Monte Carlo
air shower simulation codes. There are several parameterized atmospheric models incorporated in the CORSIKA air shower simulation code, based on averaged
profiles: U.S. standard atmosphere parameterized according to J. Linsley \cite{corsikamanual}, 
parameterized atmospheres for the Pierre Auger Observatory near Malarg\"{u}e (Argentina)
by M. Will and B. Keilhauer \cite{bianca}, South Pole atmospheres parameterized by P. Lipari and D. Chirkin etc.
So far, the US standard atmosphere has been used in LOFAR analyses, through CORSIKA simulations \cite{corsikamanual} and the CoREAS extension 
\cite{corsikamanual} which is used to calculate the radio emission of the air showers.

A first order linear correction to the US standard atmosphere has been applied to account for the fact that the 
US-standard atmosphere does not reflect the realistic atmospheric conditions at a given time. It is preferable to 
integrate a realistic atmosphere directly into the simulations. In particular, the reconstruction of $X_{\rm max}$ depends 
on the refractive index of air, and so a realistic refractive index profile needs to be included.

\textcolor{black}{The effects of the refractive index, n, on the reconstructed $X_{\rm max}$ have been previously reported in Ref.\cite{codalema} and Ref.\cite{Arthurpaper}, using
different simulation codes.
In Ref.\cite{Arthurpaper}, CoREAS was used to simulate two ensembles of showers, one with a globally
higher refractivity $N= \left(n-1\right)\, 10^6$, another with standard values. A Monte Carlo based approach was taken to study the systematic 
shift in reconstructed $X_{\rm max}$ by comparing
the set of simulations with higher refractivity to the standard ones. The shift in the reconstructed $X_{\rm max}$  from the default value 
was found to be proportional to the geometric distance to $X_{\rm max}$. The effect was stronger in the high frequency band of 120--250 MHz than in the 30--80 MHz band.
In Ref.\cite{codalema}, a more realistic profile of the refractivity was constructed
for one particular day using information from the Global Data Assimilation System, GDAS, a global weather database. The differences between this atmosphere and default atmospheres were studied
using the SELFAS radio emission simulation code \cite{selfas}. The results showed that correcting for the realistic density is the most important factor in
the accurate reconstruction
of $X_{\rm max}$, causing about 30 g/cm$^2$ bias in $X_{\rm max}$. And the second most important correction was through 
the inclusion of the high frequency refractivity formula, applicable at radio frequencies, contributing about 5 g/cm$^2$ bias in $X_{\rm max}$. The effects of the 
increased refractivity 
on the time traces and the lateral distribution function (LDF) were also
reported. In the 20--80 MHz frequency band, relatively small differences in the amplitude of the electric field and LDF were found, whereas considerable differences were 
found studying the high frequency band between 120--250 MHz. These results were in agreement with  Ref.\cite{Arthurpaper}.
While both works paved the way for the understanding 
of atmospheric effects on radio simulations, a direct application to real data using simulations with realistic atmospheric conditions was not addressed.\\
In this work, for the first time, GDAS-based atmospheric profiles, automatically included in CoREAS simulations are applied to LOFAR data. The effects of atmospheric
parameters like pressure and humidity on the reconstructed $X_{\rm max}$ are studied and compared  to the results of 
previously used linear corrections. A new GDAS-based correction is introduced and compared to previous methods. Furthermore, 
a tool is developed to extract GDAS atmospheric parameters which are then interfaced with CORSIKA. The utility of this tool is not only limited
to LOFAR. This code, called \textquoteleft{gdastool}\textquoteright, has been available for public use since the release of CORSIKA version 7.6300. 
It is flexible and ready to be adapted by the users to obtain parameterized
atmospheric profiles for user-specified time and location. Sections 2 and 3 describe the processing of GDAS data
to extract the atmospheric state variables and examples of atmospheric profiles at the LOFAR site,
respectively. Section 4 covers the details of the implementation of GDAS in CORSIKA. In sections 5 and 6, LOFAR cosmic ray data are 
evaluated with the GDAS atmospheric profiles, the GDAS-correction factor is introduced and the explicit effects of humidity on 
shower parameters are discussed. }

\section{Extracting atmospheric variables from GDAS data}
The  Global Data  Assimilation  System (GDAS) developed  at  
NOAA's\footnote{National Oceanic and Atmospheric Administration.} National Centers for Environmental Prediction (NCEP) is a
tool used to describe the global atmosphere. It is run 
four times a day (0,  6,  12,  and 18 UTC) and provides a 3-,  6- and 9-hour forecast
based on the interpolation of meteorological measurements from all over the world including 
weather stations on land, ships and airplanes as well as 
radiosondes and weather satellites~\cite{gdas}. The three hourly data are available at 
23 constant pressure levels,  from 1000 hPa (roughly sea level) to 
20 hPa ($\approx 26 \, \mathrm{km}$) on a global $1^{\circ}$ spaced
latitude-longitude grid ($180^{\circ}$ by $360^{\circ}$). Each data
set is complemented by data at the surface level. The data are stored 
in weekly files and made available online. In order to model a realistic atmosphere one
needs to obtain the suitable atmospheric
parameters from GDAS. Parameters like temperature (K),  height (m)
relative humidity ($H$) and pressure (hPa) can be directly extracted
from the database. In the GDAS data,  the altitude is in geopotential units with respect to a geoid (mean
sea level). This is an adjustment to geometric height or elevation above mean sea level using the 
variation of gravity with latitude and elevation. To convert from geopotential height $h$ (m) to
standard geometric altitude $z$ (m) we use the formula 
\begin{equation}
 z\left(h \, , \Phi\right) =  \left(1+ 0.002644 \cdot \cos(2\Phi)\right) 
 \cdot h + (1+0.0089\cdot \cos(2\Phi))  \left(\frac{h^{2}}{6245000}\right) 
\end{equation}
where $\Phi$ is the geometric latitude \cite{gdasauger}. 
To calculate the air density,  the relative humidity is to be converted
into water vapor pressure. The following approximation 
of the empirical Magnus formula is used to calculate the water vapor pressure (hPa)
in terms of humidity and temperature \cite{gdasauger}:
\begin{equation}
e = \frac{H}{100\%} \times 6.1064 \times \exp{\left(\frac{21.88\hspace{0.1cm}t}{265.5 
\hspace{0.1cm}+\hspace{0.1cm} t}\right)}  \hspace{1cm} \mathrm{for} \hspace{0.2cm}t\leq 0^{\circ}C
\nonumber
\end{equation}
and
\begin{equation}
e = \frac{H}{100\%}\times 6.1070 \times \exp{\left(\frac{17.15 \hspace{0.1 cm}t}{234.9\hspace{0.1 cm} + 
\hspace{0.1 cm}t}\right)}\hspace{1cm} \mathrm{for}\hspace{0.2cm} t\geq 0 ^{\circ}C \, .
\label{humeq}
\end{equation}

The density can be calculated from the ideal gas law as
\begin{equation}
\rho = \frac{P \hspace{0.1 cm} M_{\mathrm{air}}}{R \hspace{0.1 cm}T}
\end{equation}
where $P$ is the atmospheric pressure in Pa,  $T$ is temperature in K and $R$ is the universal gas constant, having a 
value of 8.31451 J K$^{-1}$ mol$^{-1}$ and 
$M_{air}$ is the molar mass of air. Moist air can be decomposed
into three components to calculate its molar mass: dry air, water vapor and carbon dioxide. The
molar mass of humid air is the sum of the molar masses of the components,  weighted with the
volume percentage $\phi_{i}$ of that component \cite{gdasauger}, 
\begin{equation}
M_{\mathrm{air}} = M_{\mathrm{dry}} \cdot \phi_{\mathrm{dry}} + M_{\mathrm{water}} \cdot \phi_{\mathrm{water}} + M_{\mathrm{CO_{2}}} \cdot \phi_{\mathrm{CO_{2}}} \, .
\end{equation}
The molar masses of dry air,  water vapor and CO$_{2}$ are 0.02897,  0.04401 and 0.01802~kg-mol$^{-1}$
respectively. The volume percentage of CO$_{2}$ is taken as 385 ppmv, the percentage of water $\phi_{\mathrm{water}}$ is
the partial pressure of water vapor divided by the pressure $P$; the dry air makes up the rest. \\
The refractivity, defined as $N=\left(n-1\right) \, 10^6$, is a function of humidity, pressure and temperature can be expressed as 
\begin{equation}
 N = \SI {77.6890} {\kelvin\per\hecto\pascal}\frac{p_{d}}{T} + \SI{71.2952}{\kelvin\per\hecto\pascal}\frac{p_{w}}{T} + \SI{375463}{\square\kelvin\per\hecto\pascal}\frac{p_w}{T^{2}}
 \label{ri}
\end{equation}
with $p_{w}$, $p_{d}$ and $T$ being the partial water
vapor pressure $\left(p_w= e \times 100 \, \mathrm{Pa}\right)$, partial dry air pressure and temperature respectively \cite{Rueger}. 
The effect of humidity is important for our study as it tends to increase the refractivity in comparison to that of dry air
at the radio frequencies. There are differences between the refractivities obtained in radio and the ones
in the visible, near the infrared and UV ranges as described in \cite{gdasauger}. To account for the uncertainties in GDAS data
one needs to perform in situ measurements with weather balloons. Since this is beyond the scope of this work and
we refer to \cite{gdasauger}, which provides a comparison between GDAS data and weather balloon measurements in Argentina. 
Since global atmospheric models are typically more precise in the Northern hemisphere where more
weather data is available we assume that
the intrinsic uncertainty of GDAS at the LOFAR site is
similar to that in Argentina. Various relevant uncertainties are:
$\pm$0.5 $^\circ$C for temperature, 0.5 hPa for
pressure,  and 0.05 hPa for water vapor pressure and less than 1~$\mathrm{g/cm^{2}}$ in atmospheric depth over the altitude range from 3 to 6 km. The uncertainty in water vapor
pressure translates to $2-7\%$ uncertainty in humidity. 
The resulting relative uncertainty in $N$ due to these parameters is around 0.5$\%$ at the same altitude range.
The GDAS data have a resolution of $1^{\circ}$ by $1^{\circ}$ in latitude longitude. This can be roughly approximated as a distance of 100 km
between two adjacent grid points. For highly inclined showers the distance to the region of shower development from the observation site can be larger than the distance between two grid points.
For air showers coming from  70$^{\circ}$ zenith this distance is around 70 km and for zenith $>$ 75$^{\circ}$ it is about 100 km. In these cases, the choice of
an exact grid point becomes complicated. Also at this point, for zenith angles $>$ 70$^{\circ}$ the correction due to curved atmosphere becomes important. This does not occur
for LOFAR as the detected cosmic rays are limited to within a $<$55$^{\circ}$ zenith angle due to the particle detectors used for triggering. 
In this regime the GDAS model works well.

\section{GDAS atmospheric profiles at the LOFAR site}
\begin{figure}[h!]
\includegraphics[width=0.5\linewidth]{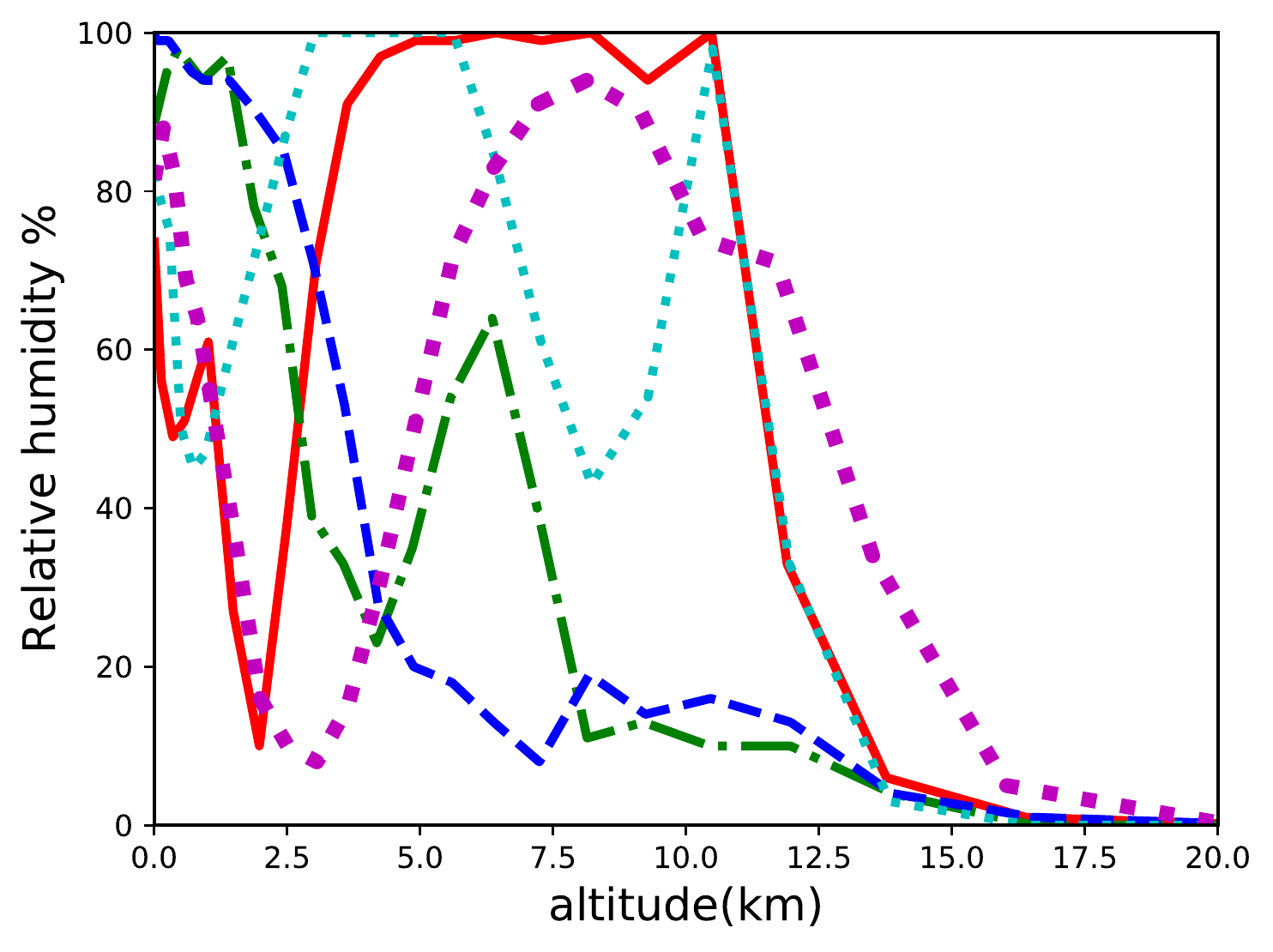}
\includegraphics[width=0.5\linewidth]{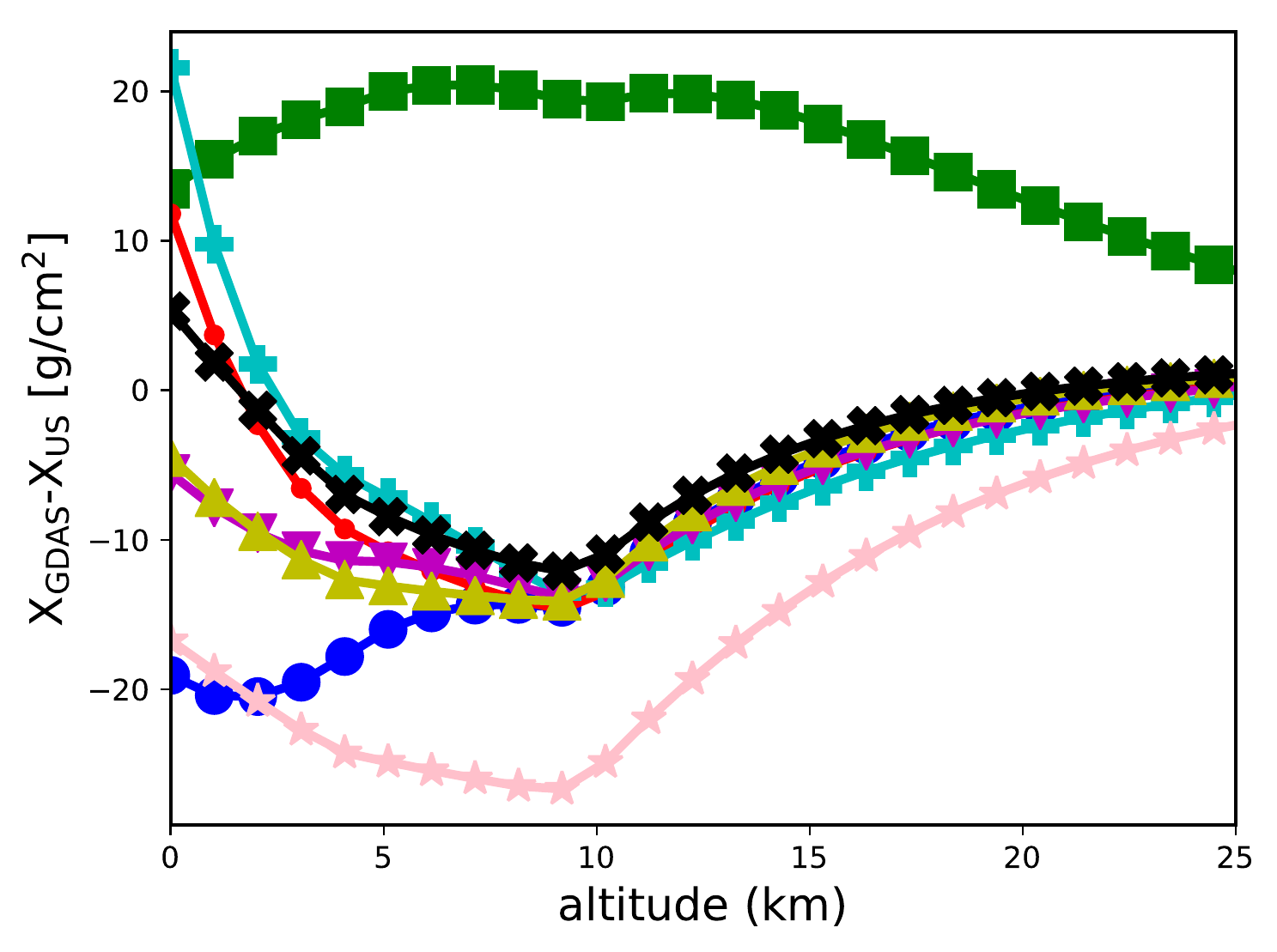}
\caption{Atmospheric profiles at LOFAR. \textbf{Left}: Example of 5 humidity profiles between June to November during the year 2011. \textbf{Right}: 8 profiles for the difference in atmospheric depth between 
  US standard atmosphere and GDAS atmospheres as a function of altitude between the years $2011-2016$. }
  \label{atm}
\includegraphics[width=0.55\linewidth]{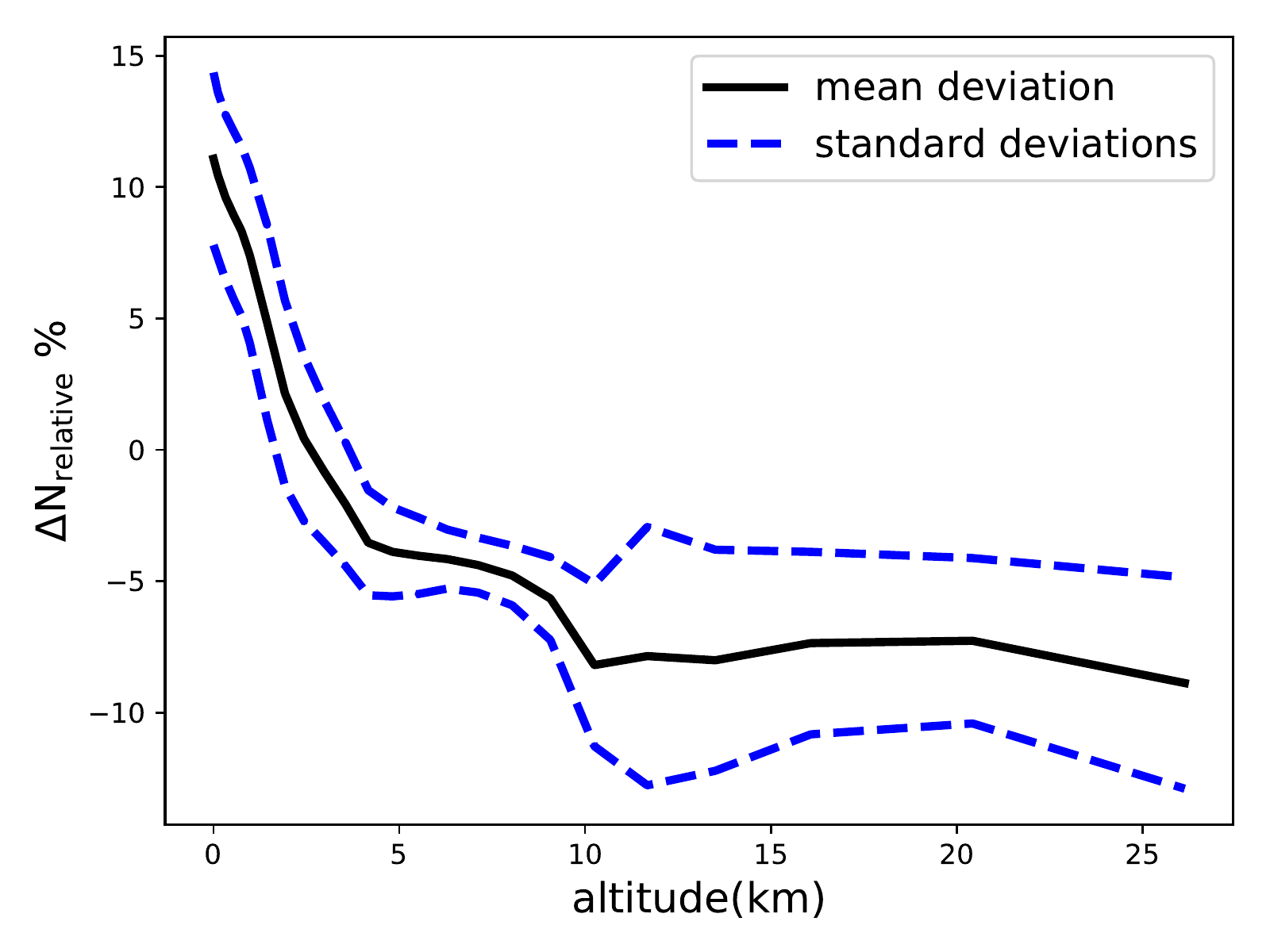}
\centering
 \caption{Mean relative refractivity, defined as $\Delta{N}_{\mathrm{relative}}= \frac{N_{\mathrm{GDAS}}-N_{\mathrm{US}}}{N_{\mathrm{US}}}$; profiles 
  for 100 recorded cosmic rays at LOFAR spanning over the years 2011 to
   2014. The black solid line denotes the mean profile and the blue dashed lines show the standard deviations. }
    \label{refrac}

\end{figure}

In this section several GDAS atmospheric profiles extracted at the LOFAR site are discussed. 
Fig-\ref{atm} (\textbf{left}) shows humidity as a function of altitude for 5 arbitrary atmospheric profiles for different
days in the year 2011, between June and November. A significant day-to-day fluctuation is seen. The red solid and blue dashed lines indicate two very different weather
conditions; the red solid line having 
high saturating humidity between $5-8$ km suggests higher cloud coverage and the blue dashed line with low humidity in that range indicates low cloud coverage. 
Fig-\ref{atm} (\textbf{right}) shows the  difference in atmospheric depth profile between the US standard atmosphere and the GDAS atmospheres at LOFAR
for 8 profiles over the years $2011-2016$. The GDAS atmospheres 
vary significantly from the US atmosphere. Atmospheric profiles with similar
atmospheric depth at ground can evolve differently higher in the atmosphere. This is important for calculating
the correct distance to the shower maximum. 
Fig-\ref{refrac} shows the mean profile for the relative difference in refractivity $\Delta{N}_{\mathrm{relative}}$ between GDAS and the US standard atmosphere as a function of altitude 
for over 3 years for 100 cosmic rays 
recorded at LOFAR. It is defined as $\Delta{N}_{\mathrm{relative}}=  (N_{\mathrm GDAS} - N_{\mathrm US})/N_{\mathrm US}$, where $N_{\mathrm{GDAS}}$ is calculated from
Eq-\ref{ri} using GDAS atmospheres at LOFAR. \textcolor{black}{$N_{\mathrm{US}}$ is obtained from the linear relation $N_{\mathrm{US}}=\frac{\rho_{\mathrm{us}}}{\rho_{\mathrm{sealevel}}} N_{\mathrm{sealevel}}$, with $N_{\mathrm{sealevel}}=292$. This is the default option for calculating refractivity in CoREAS as well}.

The absolute value of the mean $\Delta{N}_{\mathrm{relative}}$
is around $10\%$ near ground and around  $3-8 \%$ between 3 to 10 km of altitude, the region important for shower development.

Approximately 75\% of the atmospheric matter and 99\% of the total mass of water vapor and aerosols are contained within the troposphere, the lowest layer of Earth's atmosphere.
Within the troposphere the temperature drops with altitude, reaching a constant value in the tropopause, the boundary region between troposphere and stratosphere.
In the U.S standard atmosphere the troposphere ends at 11 km and tropopause extends to an altitude of 20 km. For the local GDAS atmospheres these boundaries are not sharply defined.
The flat part in the mean $\Delta{N}_{\mathrm{relative}}$ $>$ 10 km in Fig-\ref{refrac} is the result of constant temperature
in the tropopause. However contribution from this region to the radio emission is minimal.
To consider the effects of refractive index in the propagation time of radio signal it is important to calculate the effective $N$ \cite{huege2016,zharies}. This is 
defined as\\
\begin{equation}
 N_{\mathrm{eff}}=\frac{\int N(h)dh}{D}
 \nonumber
\end{equation}
where $D$ is the distance between the line of emission and observer. 
The values of relative effective refractivity $\Delta{N}_{relative}^{eff}$ between the GDAS and US standard atmosphere
are around $7-10$ $\%$ in the range of altitude mentioned above, \textcolor{black}{for observers within $<1$00 m of the shower axis.}

\section{Implementation in CORSIKA/CoREAS}
To incorporate the atmospheric parameters extracted from
GDAS in CORSIKA and CoREAS we have developed a program named \textquoteleft{gdastool}\textquoteright
\, that downloads the required GDAS file given the time and location 
of observation of the event and returns refractive indices
between ground and the highest GDAS level. It also
fits the density profile according to the standard 5 layer atmospheric
model used in CORSIKA \cite{corsikamanual}. In this model
the density $\rho(h)$ has an exponential dependence on the altitude leading
to the functional form of  mass overburden $T(h)$ which is the density 
integrated over height (km) as
\begin{equation}
T(h) = a_{i} + b_{i}   e^{-10^{5}h/c_{i}}  \hspace{1cm}i = 1, . . . , 4 \, .
\label{dep}
\end{equation}
Thus, the density is
\begin{equation}
\rho(h) = b_{i} / c_{i}   e^{-10^{5}h/c_{i}}  \hspace{1cm}i = 1, . . . , 4 \, .
\label{den}
\end{equation}
In the fifth layer the overburden is assumed to decrease linearly with height. 
The parameters $a_{i}$,  $b_{i}$ and $c_{i}$ are obtained in a manner
such that the function $T(h)$ is continuous at the layer boundaries
and can be differentiated continuously. The first three layers constitute of the 24 density points obtained from GDAS data.
\textcolor{black}{The first layer consists of 10 points, second layer of 7 points and the third layer of 7 points.}
Since GDAS provides data on constant pressure levels, not of constant heights, the layer boundaries vary slightly
between different atmospheric profiles. \textcolor{black}{The mean values of the boundaries for the conditions of 100 cosmic ray events 
are 3.56$\pm$0.11 km, 9.09$\pm$0.23 km, 26.27$\pm$0.56 km from boundary 1 to 3, respectively.}

\begin{figure}[h!]
\includegraphics[width=0.5\linewidth]{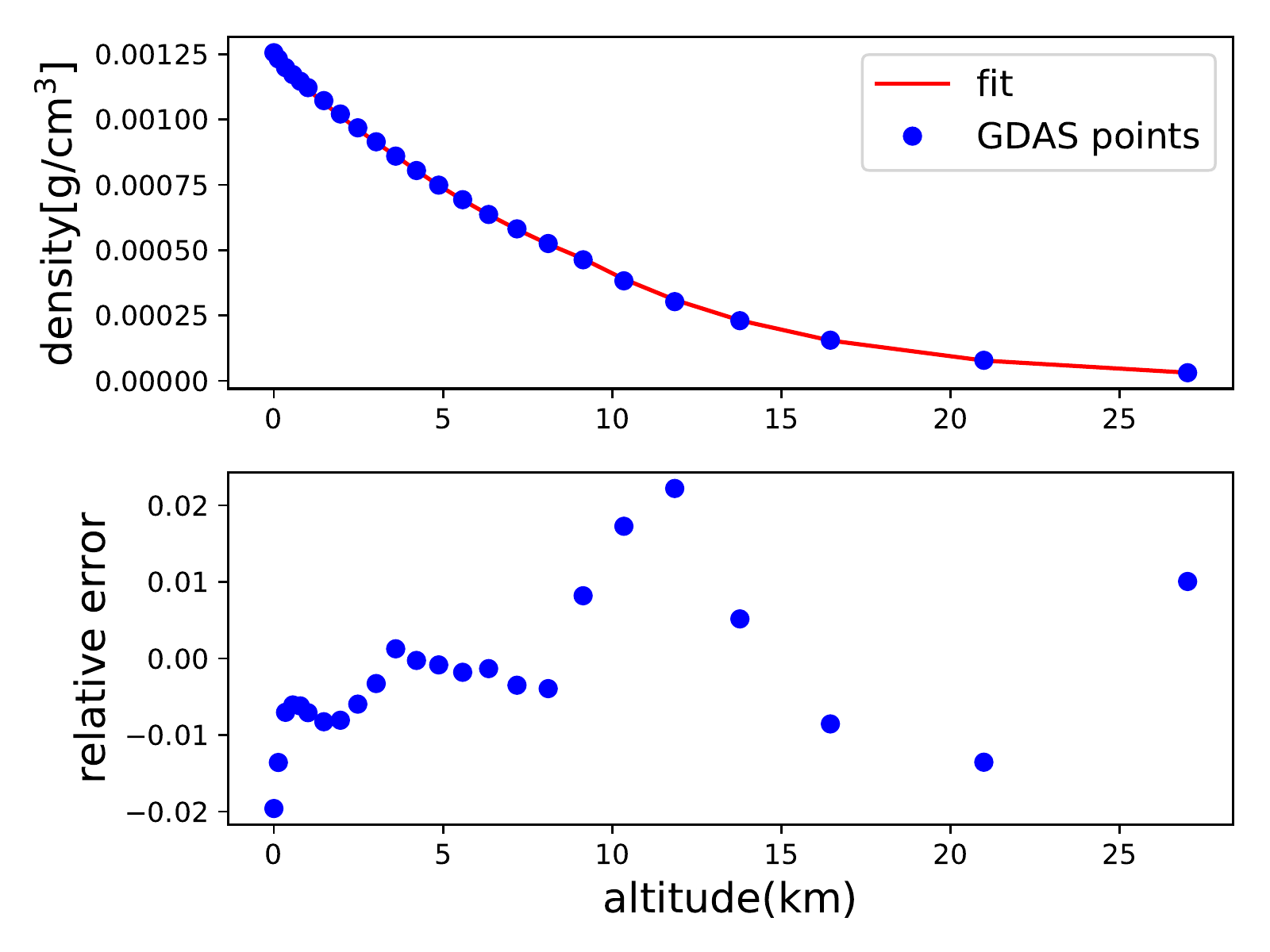}
\includegraphics[width=0.5\linewidth]{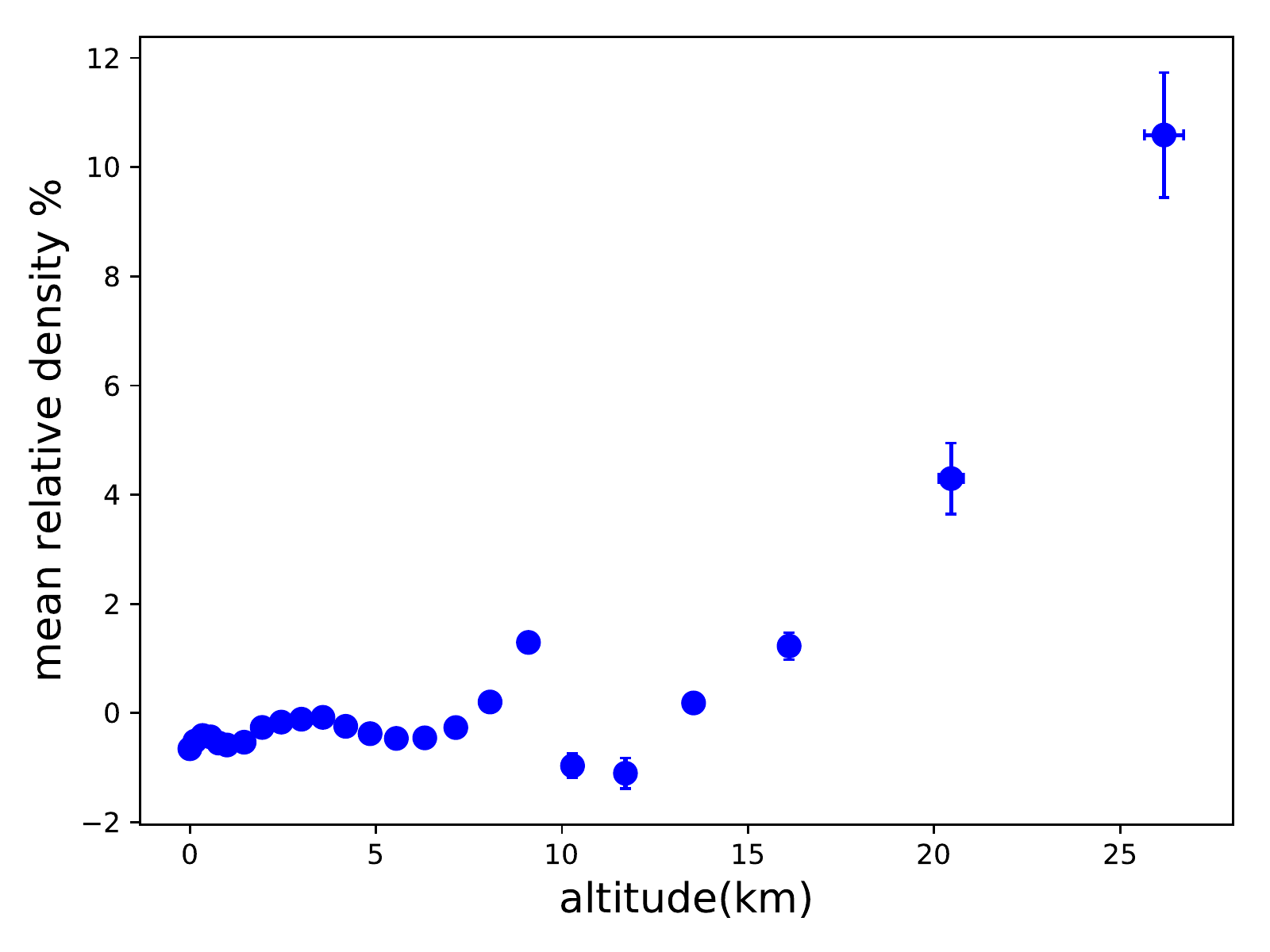}
\caption{\textbf{Left}: Example of one density profile, GDAS and the fitted 5-layered atmospheric model. The bottom
panel shows the relative error defined as $\frac{\rho_{\mathrm{fit}}-\rho_{\mathrm{data}}}{\rho_{\mathrm{fit}}}$.
\textbf{Right}: Mean relative error in density  for 100 different atmospheric profiles. The mean is calculated at each of the 24
GDAS points for all the profiles. The error bars indicate the standard deviation.}
\label{errden}
\end{figure}

Next, we fit the
data to Eq- \ref{den} in the following way:\\
For layer 1 the density profile is fitted  with two free parameters. 
Then the density $\rho_{1}$ at boundary 1 is calculated using Eq- \ref{den} with the obtained parameters $b_{1}$,  $c_{1}$. 
The condition that the density has to be continuous at the boundaries reduces
the number of free parameters to 1 which is the parameter $c$. Thus the parameter $b_{2}$ for second layer can be expressed
as a function of $\rho_{1}$ and $c_{2}$ with $c_{2}$ being the only free parameter. 
The same fitting procedure is repeated for the third layer.
\textcolor{black}{The fourth layer ranges from  the highest GDAS altitude to 100 km. At these altitudes there are no physical GDAS data. 
The parameter $c_4$ is obtained by fitting the last GDAS point and the density at 100 km from US standard atmosphere.  
At these altitudes the mass overburden is less than 0.1$\%$ of the value at ground. The important factor is to satisfy the boundary conditions throughout the atmosphere. 
Along with density the continuity of mass overburden is also preserved.}
For that, once a smooth profile for the density is obtained,
the parameter $\mathrm{a}$ in Eq- \ref{dep} is solved for analytically, using the boundary conditions for the mass overburden. 
The parameterization for the fifth layer was adapted from the US standard atmosphere \cite{corsikamanual}. 
The \textquoteleft{gdastool}\textquoteright \, also returns a density profile plot with the best fit parameters as a function of altitude and
the rms of the relative density difference between data and fit. The relative density is defined 
as $\frac{\rho_{\mathrm{fit}}-\rho_{\mathrm{data}}}{\rho_{\mathrm{fit}}}$. Fig-\ref{errden} (\textbf{left}) and its rms is used
as a goodness of fit. Fig-\ref{errden} (\textbf{left})
shows the example of a density profile between the fitted model and GDAS. 
The mean relative error in density
for 100 profiles as a function of altitude is presented in 
Fig-\ref{errden} (\textbf{right}). At lower altitudes the model fits the data very well;
deviations $>$ 2\% start at altitudes higher than 15 km which are not so important
for the shower development. A bump in the profile at 10 km is observed, this can 
be explained by the change in the atmosphere at the troposphere boundary as 
discussed in the previous section.
There will be an error on the atmospheric depth introduced by the fitted
model in Eq- \ref{dep}. It is on the order of 2 $\mathrm{g/cm ^2}$ on average between the altitude range mentioned above
with a variance of $4-5$~$\mathrm{g/cm^2}$. \\

The \textquoteleft{gdastool}\textquoteright \, can be executed as a stand alone script within CORSIKA.
Given the coordinate and UTC time stamp as input parameters it downloads the required GDAS files
and extracts atmospheric data. It then returns an output file that contains
fitted mass overburden parameters and tabulated refractive indices interpolated to 1 m intervals.
This output file can be invoked through the  CORSIKA steering file. When called, it replaces the 
default atmospheric parameters in CORSIKA with the new ones and the on-the-fly refractive index calculation
in CoREAS with the look-up table.



\section{Effects on the reconstruction of the depth of the shower maximum}
The highest precision for the determination of $X_{\rm max}$  with the radio technique
is currently achieved with the LOFAR radio telescope. Situated in the north of the 
Netherlands, the dense core of LOFAR  consists of
288 low-band dipole antennas within a circle with a diameter of 320 meters, known
as the Superterp. The radio emission from air showers in the frequency range 30--80~MHz 
is recorded by the LOFAR low-band antennas \cite{lofar,LOFAR545454}. An array of particle
detectors installed on the Superterp provides the trigger for the detection of the air showers \cite{Lora}. \\

The $X_{\rm max}$  reconstruction technique used at LOFAR is based on the production of dedicated
simulation sets for each detected air shower. The number of simulations needed to reconstruct the shower maximum
is optimized with CONEX \cite{stijnicrc}. 
A set of full CORSIKA simulations with proton and iron primaries is produced for each detected cosmic ray.
The radio emission is simulated in a star-shaped pattern for antenna positions in the shower plane using
CoREAS. An antenna model is applied to the simulated electric fields and compared to the
measured signal in the dipole antennas \cite{katiepaper}. The time integrated pulse power is calculated in
a 55 ns window centered around the pulse maximum, summed over both polarizations. Finally, a
two-dimensional map of the time integrated power is created by interpolating the star-shaped pattern \cite{xmax}. 
In the previous analysis a hybrid fitting technique was used in which  both the  radio and particle data were fitted to the 
two-dimensional radiation map and the one-dimensional particle lateral distribution function simultaneously. 
In this work  instead of the combined fit we fit only the radio data to the radio simulation. 
The advantage of switching to the radio only fitting method is that it results in reduced systematic uncertainties.



\begin{figure}[h!]
\includegraphics[width=0.5\textwidth, height=0.3\textheight]{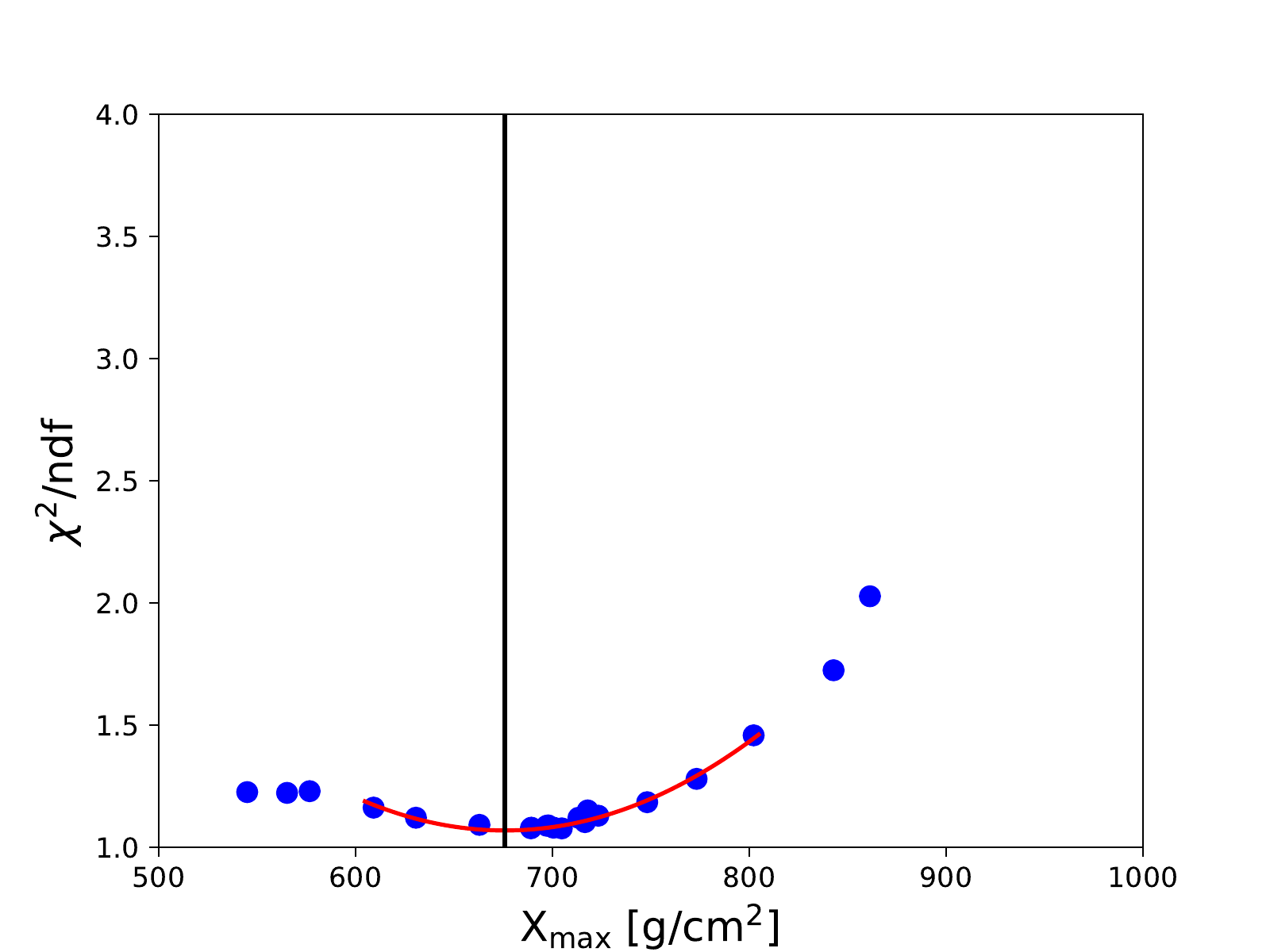}
\includegraphics[width=0.5\textwidth, height=0.3\textheight]{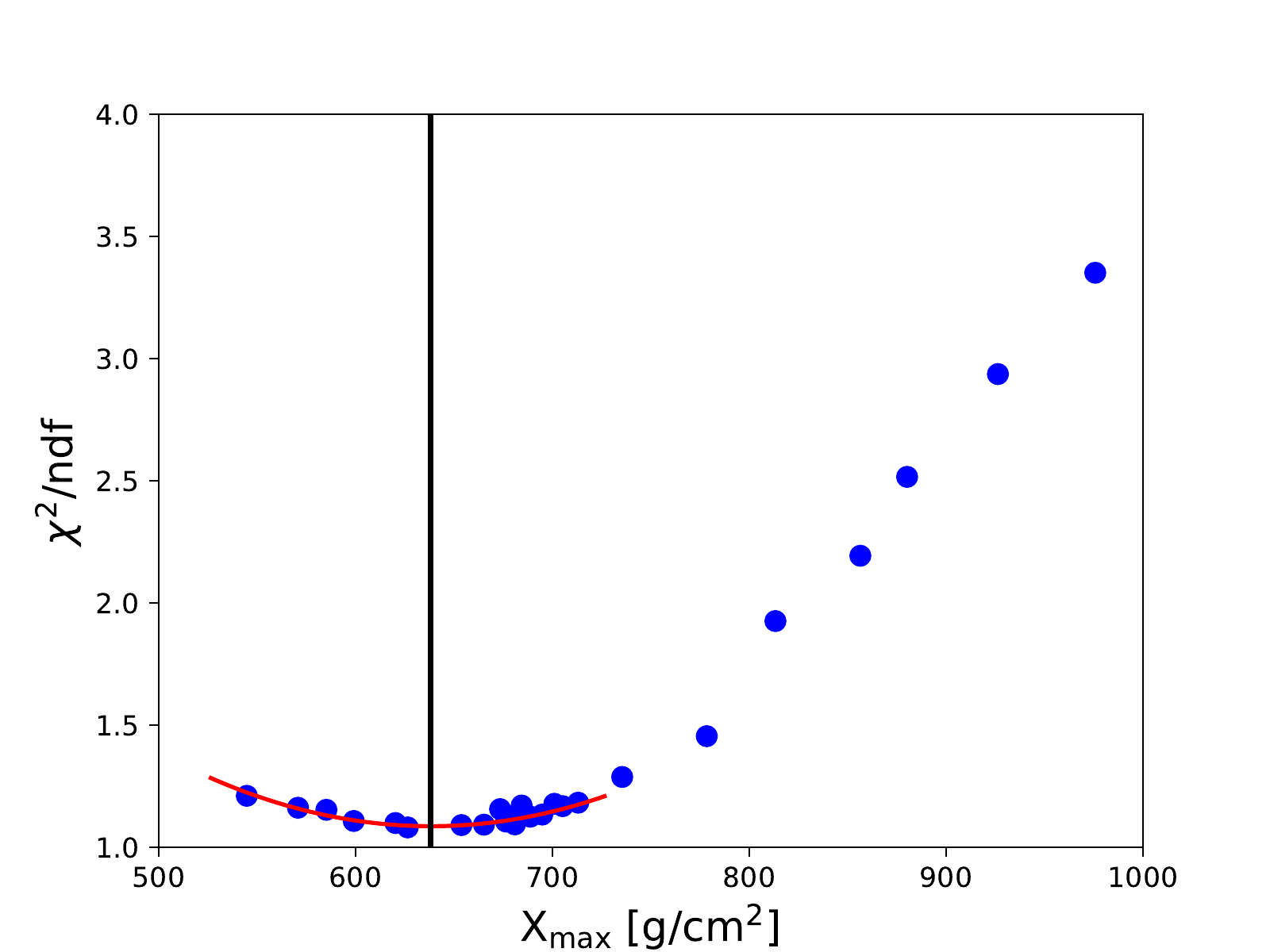}
\caption{Quality of fit as a function of simulated $X_{\rm max}$  for a LOFAR event of
energy $\mathrm{1. 4\times 10^{8}}$ GeV, with a zenith angle of 38$^{\circ}$. \textbf{Left}: simulated with default US standard atmosphere,
reconstructed $\mathrm{X_{\max}}=$ 675.8
$\mathrm{g/cm^{2}}$. Applying the linear first order atmospheric correction, the resulting $\mathrm{X_{\max}}=$ 658 $\mathrm{g/cm^{2}}$. \textbf{Right}:
simulated with GDAS atmosphere, reconstructed $\mathrm{X_{\max}}=$ 638.3~$\mathrm{g/cm^{2}}$, 
the reconstructed $X_{\rm max}$  in both the cases is indicated by solid black lines. }
\label{fitdemo}
\end{figure}

Fig-\ref{fitdemo} shows the fit quality for an air shower detected with LOFAR as a function of  $X_{\rm max}$  simulated with
two different atmospheres - one with the corresponding GDAS atmosphere and the other with the US standard atmosphere. The reconstructed 
value of $X_{\rm max}$  is found from the minimum of the fitted parabola around the best fitted points. 
We chose a LOFAR event for which the ground pressure was much lower than the US standard atmosphere, by 20 hPa. The atmospheric profile for this 
particular event is represented by the blue line with circles in Fig-\ref{atm} (\textbf{right}).
The reconstructed $X_{\rm max}$ with the US atmosphere corresponds to a much higher mass overburden
than the reconstructed $X_{\rm max}$using much thinner GDAS atmosphere.
In this example this translates to a difference of around 37.5~$\mathrm{g/cm^{2}}$
in the reconstructed $X_{\rm max}$  between the two cases. This large deviation is attributed to the
extreme weather condition for the shower chosen in the example. In the previous LOFAR analysis a correction
factor to the US atmosphere was used to account for the real atmosphere \cite{nature,xmax}.
The simulations that are produced with US standard atmosphere would approximately yield the correct geometrical altitude to the shower maximum. Then the 
corrected $X_{\rm max}$  is calculated by integrating the GDAS density profile obtained at LOFAR, from the top of the atmosphere
to the geometric altitude of $X_{\rm max}$  in the following way:\\
\begin{equation}
X(h)= \frac{1}{\cos\theta} \int_{h}^{\infty}\rho_{\mathrm{gdas}}(h) dh \, .
\end{equation}
The corrected $X_{\rm max}$  for this particular example is 658 $\mathrm{g/cm^{2}}$ and the difference between the
corrected and new $X_{\rm max}$  is about 20 $\mathrm{g/cm^{2}}$.\\

\begin{figure}[h!]
\begin{center}
\includegraphics[width=0.7\textwidth]{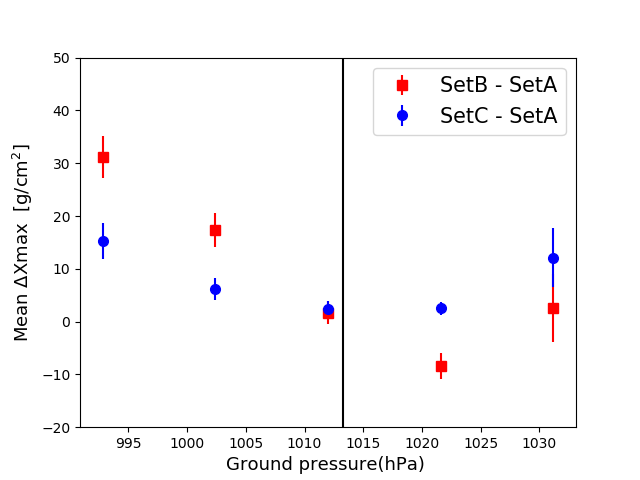}
\caption{Difference in mean $X_{\rm max}$  as a function of ground pressure. The total sample contains
123 air showers recorded at LOFAR. The black line denotes the U.S standard atmospheric pressure.  }
\label{Pplot}
\end{center}
\end{figure}
Using the same approach described above we have studied 123 air showers recorded with LOFAR with three simulation sets:

\begin{itemize}
 \item \textbf{Set A}\textendash the showers were simulated with CORSIKA v-7.6300 and GDAS atmosphere. 
 \item \textbf{Set B}\textendash the showers were simulated with CORSIKA v-7.4385 and US standard atmosphere. 
 \item \textbf{Set C}\textendash this set is identical to \textbf{Set B} but with the additional atmospheric correction factor to it as described above.
 
\end{itemize}
The effect of using  different CORSIKA versions on the reconstructed $X_{\rm max}$ , irrespective of the atmospheric model,
was probed. The difference in $X_{\rm max}$ found using CORSIKA versions 7.6300 and 7.4385 was found to be very small, around 1.4 $\mathrm{g/cm^{2}}$.
This confirms that the differences between Set-A, Set-B and Set-C are due to different atmospheric models, not any artifact arising from different versions of CORSIKA.

In Fig-\ref{Pplot} the difference in mean reconstructed $X_{\rm max}$  between the various simulation sets mentioned above is plotted against
ground pressure bins obtained from GDAS. Both the blue circles and red squares converge to zero where GDAS pressure approaches the US standard pressure
at 1013 hPa. 
The red squares have large $\Delta\mathrm{X_{\max}}$ in general. This is expected as there is no atmospheric correction involved in Set-B. 
The blue circles  however show a higher deviation both at low and high  pressure values. This suggests that the linear first order correction added
to the standard US atmosphere implemented in Set-C is not sufficient. As the refractive index effects can not be included in the linear first order correction, 
one needs full GDAS-based atmospheric profiles for more extreme atmospheric
conditions.\\ 
\begin{figure}[h!]
\includegraphics[width=0.5\textwidth, height=0.3\textheight]{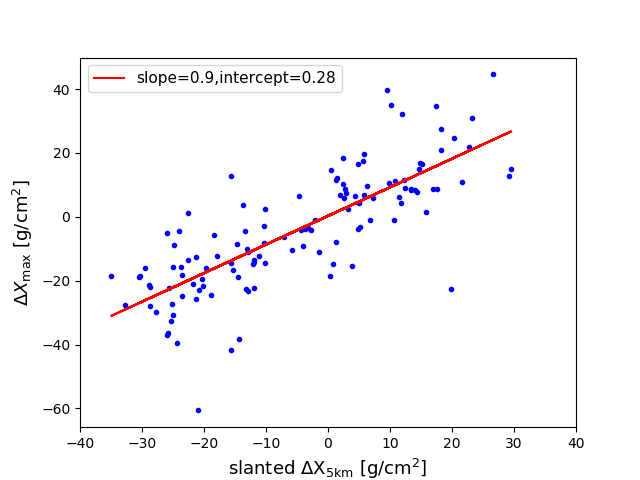}
\includegraphics[width=0.5\textwidth, height=0.3\textheight]{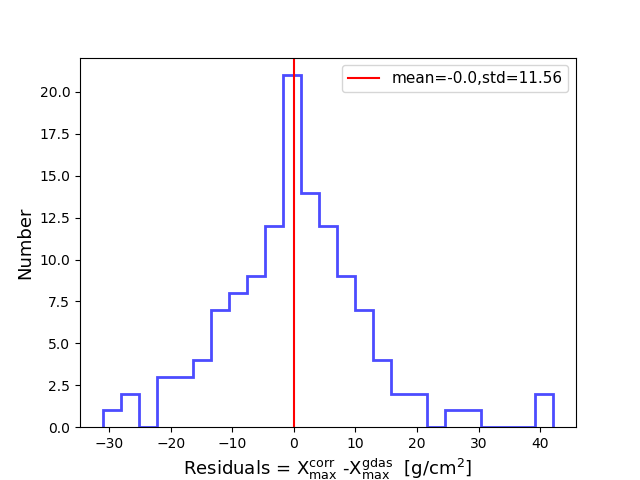}
\caption{\textbf{Left}: scatter plot of $\Delta{X_{\rm max}}= X_{\rm max}^{\rm gdas} - {X_{\rm max}^{\rm us}}$ vs difference in slanted mass overburden  
$\Delta{X_{\rm 5km}}= X_{\rm 5km}^{\rm gdas} - {X_{\rm 5km}^{\rm us}}$. The red line is a linear fit to the profile. \textbf{Right}: Histogram shows the 
residual of fitted and actual $X_{\rm max}$; residual= $X_{\rm max}^{\rm corr} - X_{\rm max}^{\rm gdas}$.}
\label{globalcorr}
\end{figure}

\noindent \textcolor{black} {Here, we study the possibility to introduce a new global correction factor to the reconstructed $X_{\rm max}$ with US standard atmosphere to correct 
for realistic atmospsheres without having to run full GDAS-based CoREAS simulations. To achieve this we studied the correlation between $X_{\rm max}$, 
refractivity, and slanted mass overburden which is defined as the integrated density from the edge of the atmosphere to a given height
at the slant of zenith angle, at different altitudes. 
It was seen that both the correlation between $X_{\rm max}$ and refractivity and between $X_{\rm max}$ and slanted mass overburden correlation   
are poor at ground and at lower altitudes. At the higher altitudes, between 4 - 6 km,
$X_{\rm max}$ and mass overburden show a higher correlation which is not prominent in $X_{\rm max}$ vs refractivity profiles at these altitudes. We have found the
strongest correlation at an altitude of 5 km.
Fig-\ref{globalcorr} (left) shows the scatter plot of $\Delta{X_{\rm max}}$ defined as $X_{\rm max}^{\rm gdas} - {X_{\rm max}^{\rm us}}$ and difference in
the slanted mass overburden $\Delta{X_{\rm 5km}}= X_{\rm 5km}^{\rm gdas} - {X_{\rm 5km}^{\rm us}}$. The precise correlation suggests 
the profile can be fit with a straight line and is used as a parameterization of global correction factor, provided by the equation:
\begin{equation}
 X_{\rm max}^{\rm corr} -  X_{\rm max}^{\rm us} = 0.9\left(X_{\rm 5km}^{\rm us}-X^{\rm gdas}_{\rm 5km}\right) +0.28 .
\end{equation}
The histogram in Fig-\ref{globalcorr} (right) shows the residual of the  $X_{\rm max}^{\rm corr}$
from ${X_{\rm max}^{\rm gdas}}$. The profile is symmetric with mean 0 g/cm$^2$ and standard deviation 11.56 g/cm$^2$. The fluctuations are within the typical 
systematic uncertainty of the reconstructed $X_{\rm max}$ with LOFAR, 
which is around 17 g/cm$^2$ \cite{xmax}. This correction factor can be used as a rule of thumb for the estimation of 
reconstructed ${X_{\rm max}}$  with the following caveats. It is specific to LOFAR, as simulations  were performed involving weather conditions, observation level, 
and magnetic field particular to LOFAR. Corresponding correction equations for other experiments can be constructed in the same manner and
can yield different results depending on atmospheric parameters. \\
However, while this global correction factor is very useful when a fast reconstruction is needed, we will use the full Monte Carlo approach in a future composition analysis. 
Simulations with event specific GDAS atmospheres are always more
accurate than the correction factor. The correction factor might also introduce biases related to the mass of the primary particles.
Proton primaries on average generate showers that reach maximum lower in the atmosphere than iron; these kind of effects are not taken into account.}



\section{Effects of humidity}

\begin{figure}[h!]
\begin{center}
\includegraphics[width=0.65\linewidth]{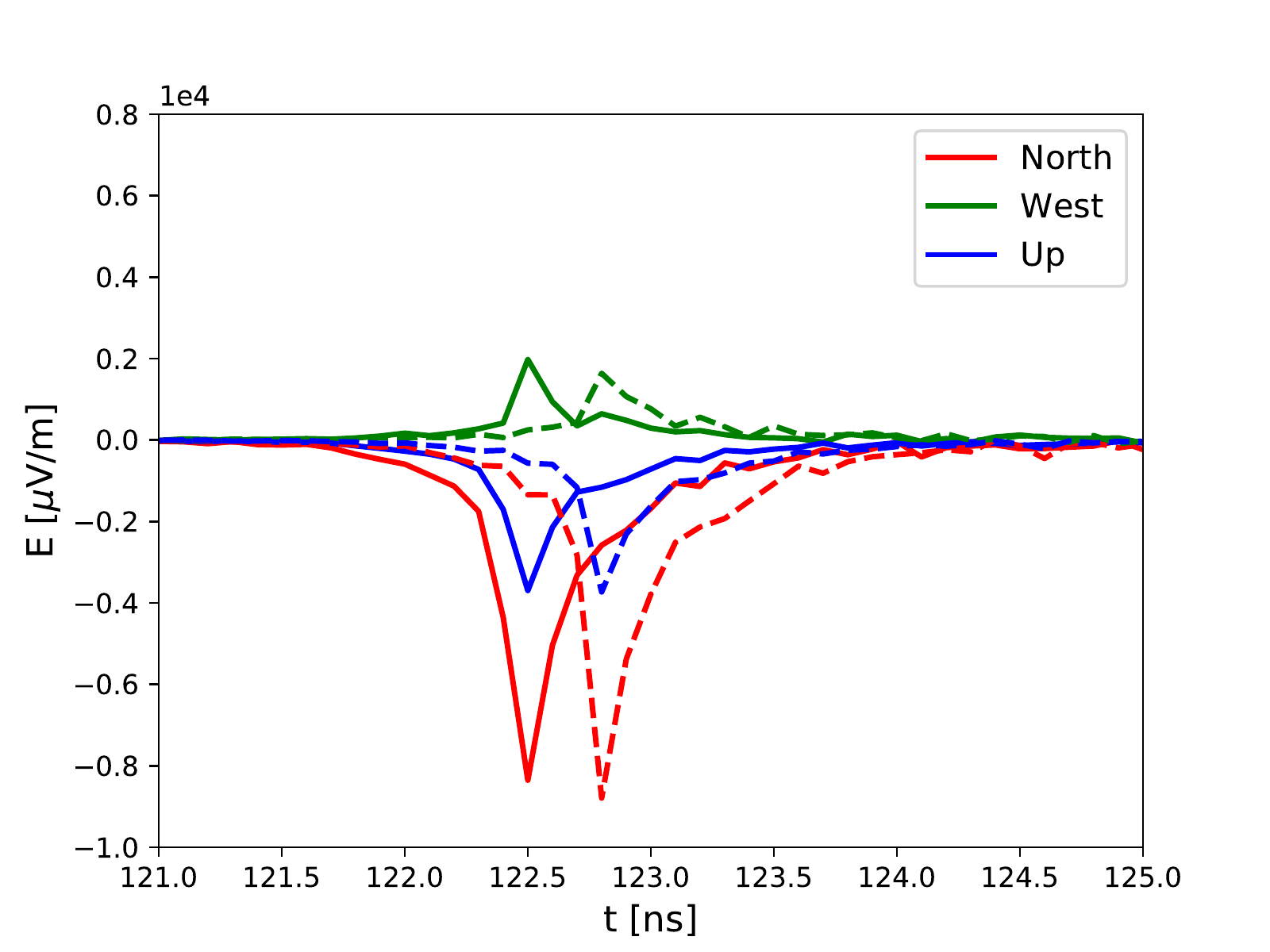}
\caption{Unfiltered electric field components of a CoREAS pulse
in time for two different refractive index profiles for a
10$^{17}$ eV proton shower with a zenith angle of 45$^\circ$ coming from east for an observer at 150 m from the axis. The solid and dashed lines 
represent the profiles with lower and higher refractive indices respectively. }
\label{time}
\end{center}
\end{figure}

\begin{figure}[h!]
\includegraphics[width=0.5\linewidth]{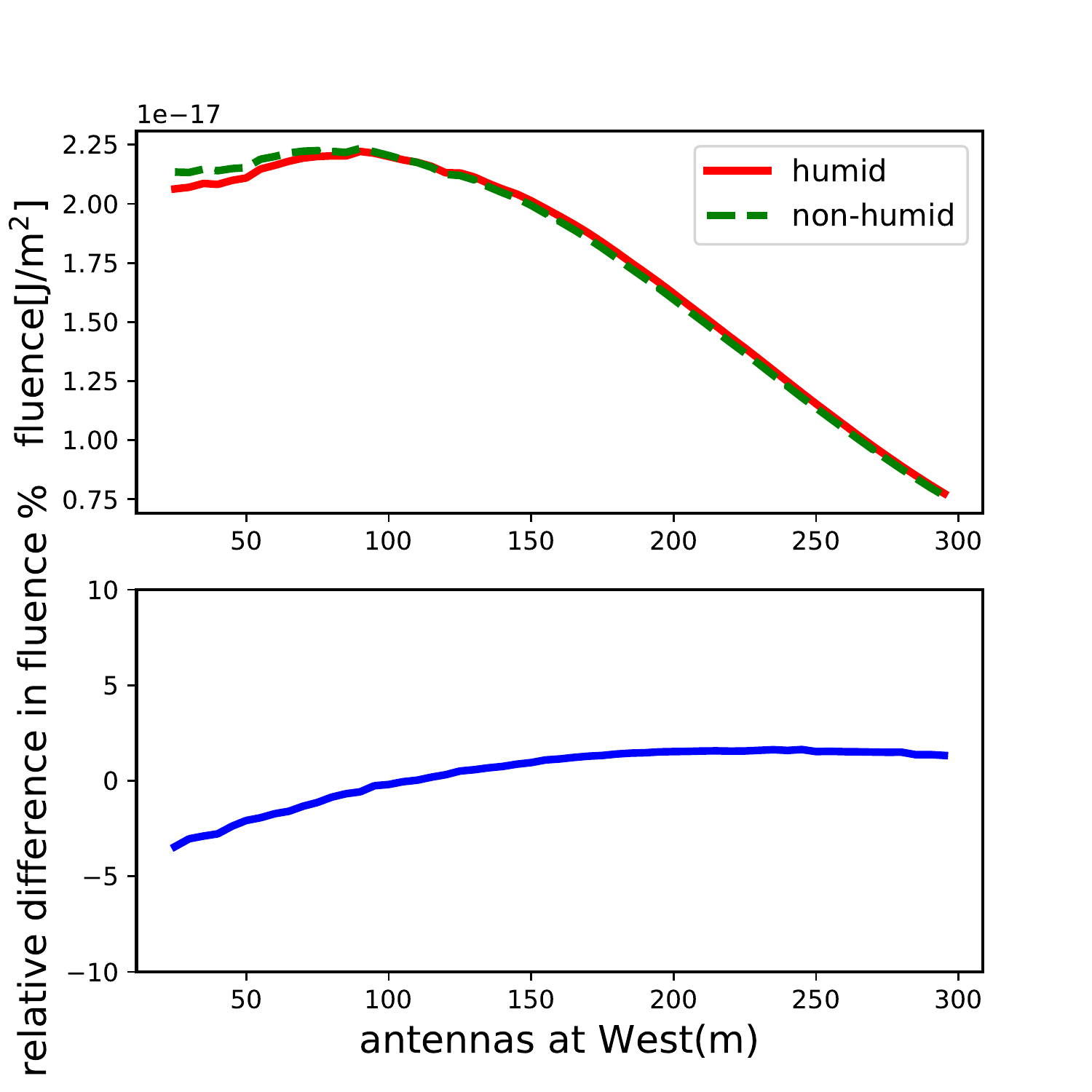}
\includegraphics[width=0.5\linewidth]{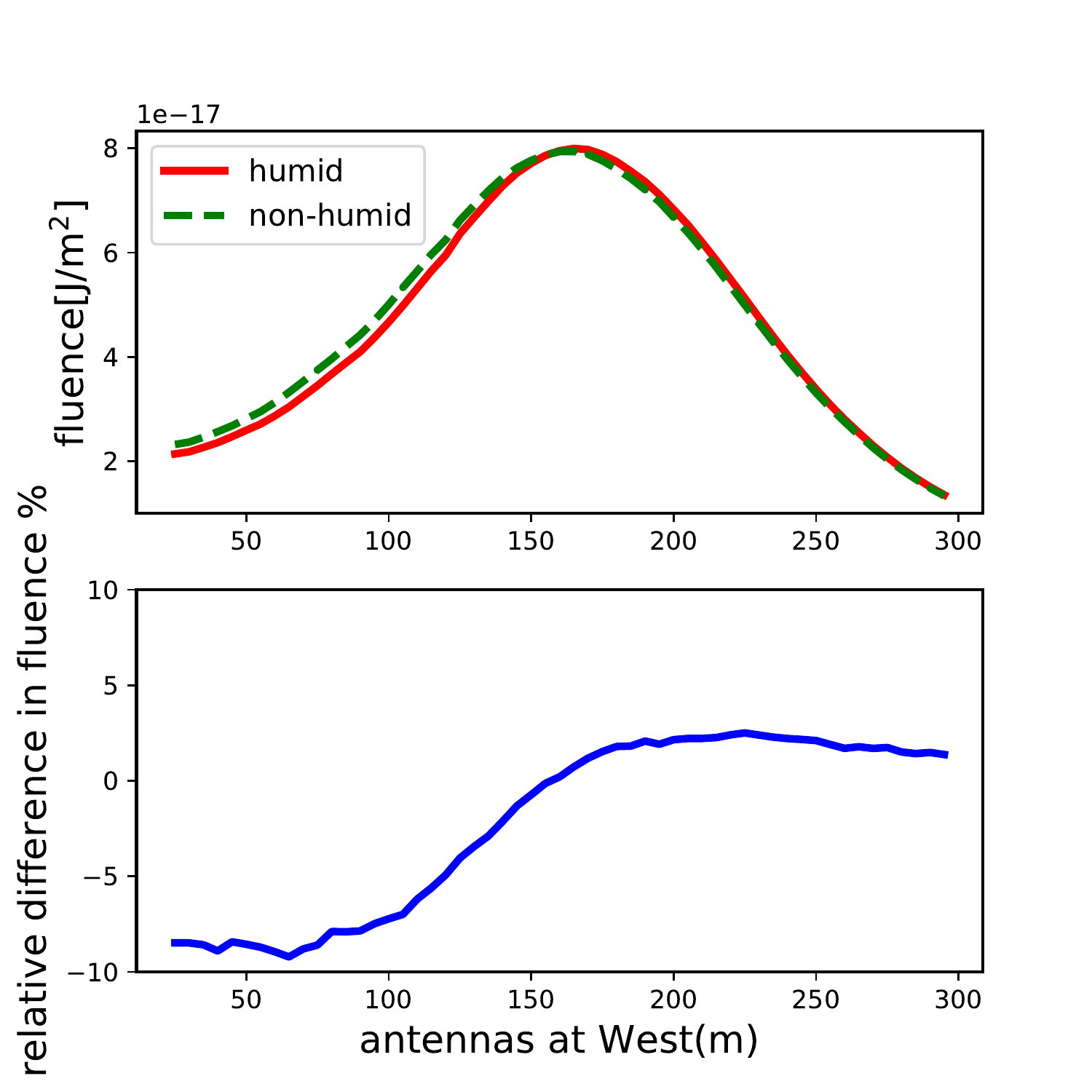}
\caption{LDF profiles for a $10^{17}$ eV proton shower coming from zenith 45$^\circ$ with $\mathrm{X_{\max}}=~593\mathrm{g/cm^{2}}$. Observers are located
to the west of the shower axis. \textbf{Left}: low frequency band between 30--80~MHz, \textbf{Right}: high frequency band between
50--350~MHz. The upper panel shows the LDF of total fluence for the humid and non-humid sets,  the lower panel shows the relative difference between these two. }
\label{cheren}
\end{figure}

As described in section 2, in the radio frequency regime, humidity increases the
refractive index. For this study, two sets of simulations were produced. In one set the showers were simulated with the respective GDAS atmosphere and in the other
with a GDAS atmosphere with vanishing humidity. This was achieved
by hard-coding the partial water vapor pressure in Eq-\ref{humeq} to negligible values. 
For the GDAS atmosphere an extremely humid weather condition at the LOFAR site was chosen. 
The same atmospheric
parameters are used in both cases to ensure that the particles evolve in a similar way in the atmosphere and produce same
shower maximum. In this way the inclusion of humidity only influences the simulated radio pulses.
The difference in the refractive index  manifests in terms of propagation effects on the pulse arrival time
and power. The pulse propagating though an atmosphere
with higher refractive index will have a lower velocity compared to dry air. This results in a delayed arrival time of the signal, as seen in Fig-\ref{time}. 
The difference in peak arrival time is less than 1 ns for an observer at 150 m. The effect is found to be less prominent
for observers further away from the axis. The lateral distribution of the energy fluence, the time-integrated power per unit area, for different observer
positions is also studied for different frequency bands for these two cases, as shown in Fig-\ref{cheren}. In the low frequency band of 30--80~MHz
relevant for LOFAR the difference in the fluence between the two sets is small, \textcolor{black}{from around 4$\%$ closer to shower axis to 2$\%$ at a distance of 100~m from the axis. }
In the high frequency band of 50--350~MHz the
values are larger, being around 8$\%$ at 100 m from the core. In the higher frequency band the Cherenkov-like effects
become stronger and the signal is compressed along the Cherenkov ring \cite{nelles2015}. 
A rough estimate of the radius of the ring can be obtained from the projection of a cone with
an opening angle given by the Cherenkov angle starting
from the shower maximum. The opening angle is strongly dependent on the index of refraction. 
This explains the higher difference in power in Fig-\ref{cheren}. \textcolor{black}{ 
Similar effects in high and low frequency bands were also reported in \cite{codalema} by studying the LDF of the electric field profiles.}
Inside the Cherenkov radius pulses 
are stretched due to refractive index effects. For higher refractive
indices this will lead to lower pulse power which explains the negative
sign in the relative fluence for observer distances close to the core.


\begin{figure}[h!]
\includegraphics[width=0.5\textwidth, height=0.3\textheight]{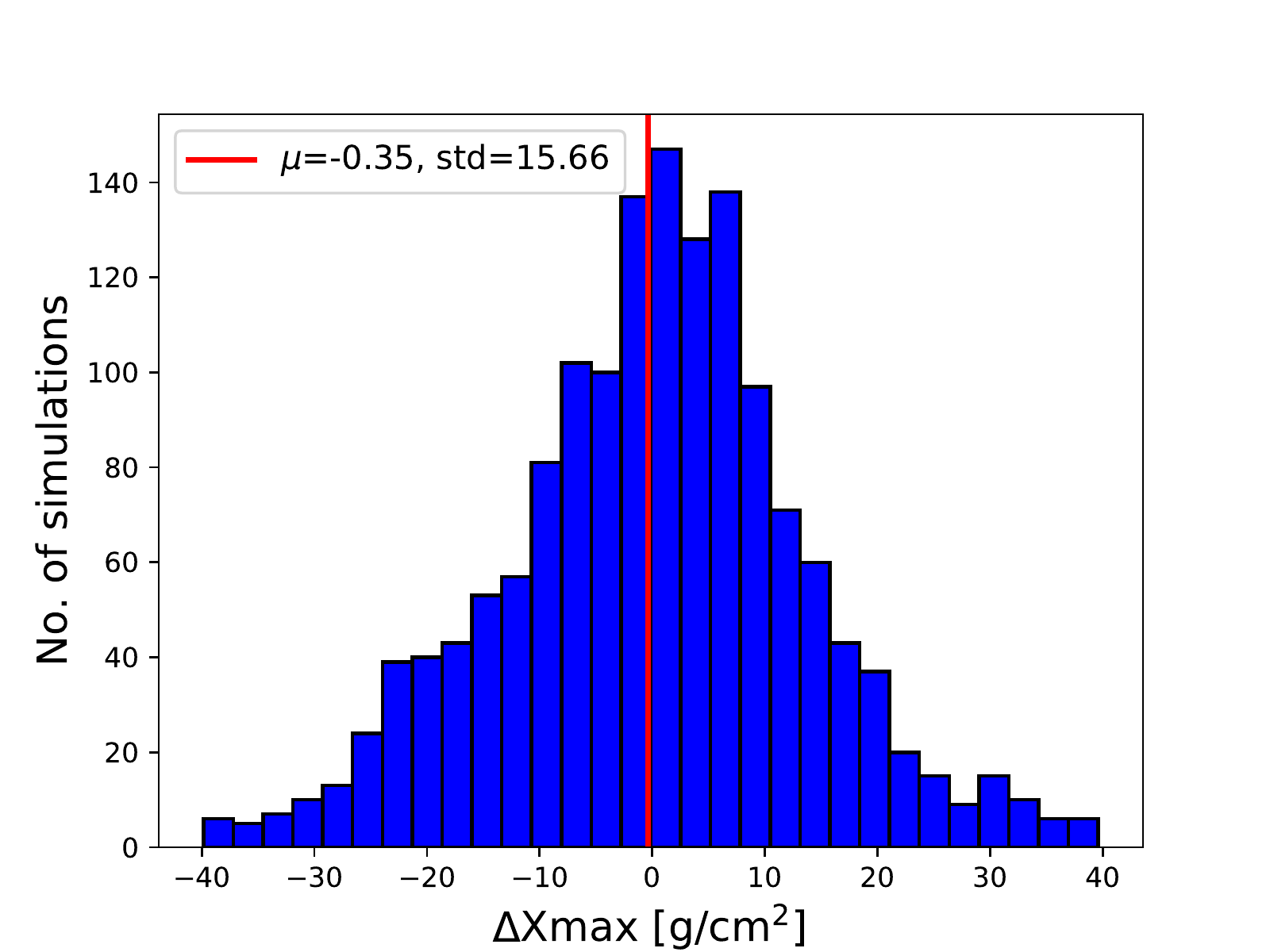}
\includegraphics[width=0.5\textwidth, height=0.3\textheight]{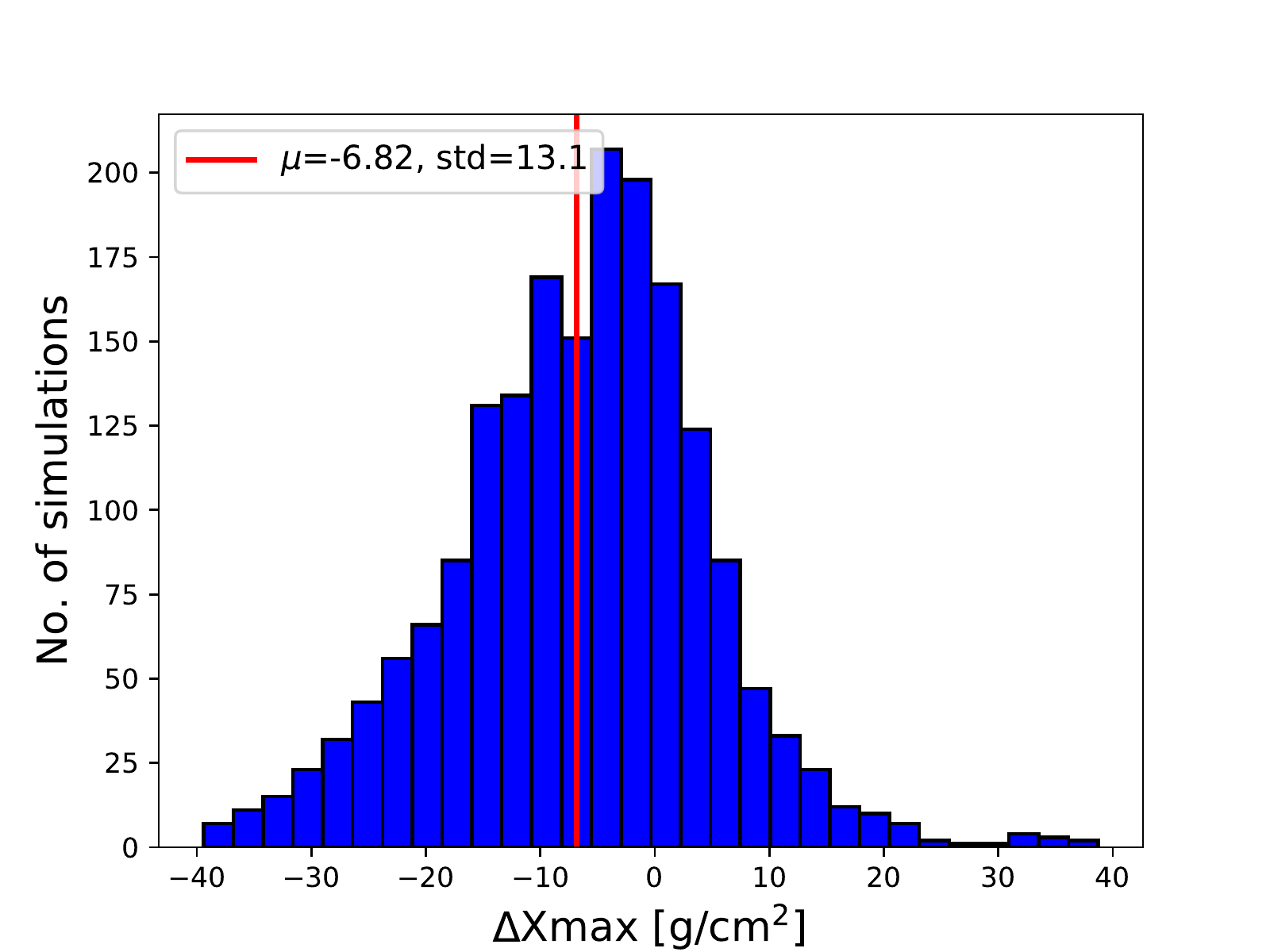}
\caption{Histogram for the $\mathrm{\Delta{X_{\max}}}=\mathrm{X_{reco}} -\mathrm{X_{real}}$ between the reconstructed and true 
value of the $X_{\rm max}$  obtained from the Monte Carlo study between the humid and non-humid simulation sets. \textbf{Left}: for the low frequency band of 30--80~MHz. 
\textbf{Right}: for the high frequency band of 50--350~MHz. The shift in the $X_{\rm max}$  is significant at 2$\sigma$ level. }
\label{histxmax}
\end{figure}
The radiation energy is the total energy contained in the radio signal. It scales quadratically with the cosmic ray energy, thus can
be used as a cosmic ray energy estimator \cite{Augerrad,radioprl}. The surface integral over the radio LDF mentioned above yields
the radiation energy. The relative difference in the integrated LDF between the humid and non-humid profiles for both the low and
high frequency regimes is smaller than 1$\%$. This indicates that humidity has almost no effect on the estimated cosmic ray energy
as determined from the radiation energy which was also concluded in \cite{Glaser}.

Next, to investigate the effect of humidity on $X_{\rm max}$  measurements we have performed a Monte Carlo comparison study between two sets of simulations that
deals with the atmospheres in a similar way as described in the beginning of this section. 
\textcolor{black}{For each of theses cases we have used a set of 40 simulated events with different energy, zenith and azimuth angles. Each of these sets consist of an ensemble of proton and iron initiated showers based on CONEX selection criteria.} 
One shower from the set with higher humidity is taken as reference and all the simulated showers from the set with zero humidity are used to perform the reconstruction. This yields a reconstructed
$\mathrm{X_{reco}}$ that can be compared to the actual $\mathrm{X_{real}}$ of the reference shower. 
The same method is repeated for all the showers in the set with higher humidity. \textcolor{black}{Showers with extreme values of $X_{\rm max}$  were not included in the fit. The range of the fit 
was taken as $\pm$ 50 $\mathrm{g/cm^{2}}$ of the actual $X_{\rm max}$  for the test shower.}

The difference $\mathrm{X_{reco}} -\mathrm{X_{real}}$ estimates the effect of humidity on the reconstructed $X_{\rm max}$. 
We do not observe any significant shift in $X_{\rm max}$  in this study. 
This indicates that these effects are most likely smaller
than the overall resolution in reconstructed $X_{\rm max}$  in the LOFAR frequency band. We also performed the same study in a higher frequency band between 50 and 350 MHz, corresponding to the SKA-low band. There, an overall shift of 6.8 $\mathrm{g/cm^{2}}$ in the reconstructed $X_{\rm max}$  was observed. These results, shown in Fig-\ref{histxmax}, are in line with the LDF studies described
earlier in this section. \\
\textcolor{black}{In Ref.\cite{Arthurpaper}, larger shifts of about 10 to 22 g/cm$^2$ in reconstructed $X_{\rm max}$  in the high frequency band of 120--250~MHz
for  4\% higher refractivity and 3.5 to 11 g/cm$^2$ in the low frequency band of 30--80~MHz were reported. A toy model was used to describe the effects. The toy model was based
on the assumptions that the size of the radio footprint on the ground would be proportional to the geometric distance to $X_{\rm max}$ and to the Cherenkov angle
at the altitude of $X_{\rm max}$. The effect of constant higher refractivity would  correspond to a higher Cherenkov angle resulting in an underestimation of $X_{\rm max}$.
This then leads to a clear linear relation between shift in $X_{\rm max}$ and distance to $X_{\rm max}$.
Without having prior knowledge of individual atmospheric conditions, an overall scaling of the refractivity profile had to suffice. However,
the realistic scenario is quite different. There are strong interplays between humidity, pressure, and temperature which are reflected in refractivity. 
The relative refractivity profile in Fig-\ref{refrac} shows that
the shift is not a constant, but is altitude dependent. From near ground to higher altitudes it switches from being a higher value than US standard atmosphere to a lower value.
This makes an one-to-one comparison to Ref.\cite{Arthurpaper} hard. However, we can argue that qualitatively same trait in the high and low frequency band has been found in 
both the works.}

\textcolor{black}{The effects of different zenith angles, true $X_{\rm max}$  and energy  were probed for the shift in $X_{\rm max}$  for both the frequency bins. The simulation set was divided in two groups, each group belonging to high and low values of the parameters mentioned above. No significant effect was seen. }


\begin{table}[h!]
\begin{center}
\begin{tabular}{ c c c }
 Frequency band & Zenith & $\Delta\mathrm{X_{max}}$ ($\mathrm{g/cm^2}$) \\ 
 50--350 MHz & low $< 30^\circ$ & -6.24$\pm$0.30 \\
 50--350 MHz & high $>30^\circ$  & -6.19$\pm$ 0.37 \\  
30--80 MHz & low $< 30^\circ$ & 0.10$\pm$0.50 \\
30--80 MHz  & high $>30^\circ$ & -0.05$\pm$0.46 \\
\hline
\end{tabular}

\begin{tabular}{ c c c }
 Frequency band & True $X_{\rm max}$  ($\mathrm{g/cm^2)}$ & $\Delta\mathrm{X_{max}}$ $(\mathrm{g/cm^2}$) \\ 
 50--350 MHz & low $<624$ & -6.78$\pm$0.41 \\
 50--350 MHz & high $>624$ & -6.30$\pm$ 0.32 \\  
30--80 MHz & low $<624$ & -0.61$\pm$0.51 \\
30--80 MHz & high $>624$ & 0.51$\pm$0.46 \\
\hline
\end{tabular}

\begin{tabular}{ c c c }
 Frequency band &  Energy(GeV)  & $\Delta\mathrm{X_{max}}$ ($\mathrm{g/cm^2}$) \\ 
 50--350 MHz & low $ <2.18\times 10^8$ & -6.86$\pm$0.35 \\
 50--350 MHz & high $ >2.18\times 10^8$  & -6.92$\pm$ 0.38 \\  
30--80 MHz & low $ <2.18\times 10^8$& -0.48$\pm$0.48 \\
30--80 MHz & high $ >2.18 \times 10^8$& 0.$\pm$0.49 \\
\hline
\end{tabular}
\caption{Shift in $X_{\rm max}$  for different zenith, energy and  $X_{\rm max}$ 
bins for different frequency bands.\label{mytable}}
\end{center}
\end{table}

\section{Conclusion and discussion}
\textcolor{black}{Simulating  air showers with realistic atmospheres is important for the precise
reconstruction of $X_{\rm max}$ with the radio technique. The GDAS database is a useful platform to extract atmospheric parameters for a given time and location.
Atmospheric effects on radio simulations were 
previously studied in Refs. \cite{Arthurpaper} and \cite{codalema}. The studies demonstrated the role of correct description of atmospheric density and refractive index
when included in the radio simulation codes. However, the application of simulations with realistic atmospheres to real data was not
addressed.  \\}
\textcolor{black}{We report, for the first time, the application of GDAS-based atmospheric profiles, automated in CoREAS
simulation to cosmic ray data. By systematically performing GDAS-based CoREAS simulations for the LOFAR dataset, we have done comparison between 
GDAS-based atmospheres a linear geometrical first order correction to the US standard atmosphere on $X_{\rm max}$. While the linear correction is sufficient for 
the bulk of the events, it becomes indispensable to use full GDAS based atmospheres for extreme values of the air pressure.
When the air pressure at ground level differs by less than 10 hPa from the US standard atmosphere value, the reconstructed $X_{\rm max}$  value 
including the linear correction agrees with the full GDAS-based reconstruction value within 2 $\mathrm{g/cm^{2}}$. However, when the ground pressure is more than 
10 hPa from the US standard atmosphere, this difference grows significantly up to 15 $\mathrm{g/cm^{2}}$. }\\

\textcolor{black}{We have also introduced a GDAS-based correction factor for $X_{\rm max}$ reconstructed with US standard atmosphere without having to run full GDAS-based CoREAS simulations. 
It is specific to LOFAR, but similar relations can be worked out for other experiments as well. The uncertainty on the predicted $X_{\rm max}$ using the correction 
factor is about 12 g/cm$^2$; this is within the typical $X_{\rm max}$ reconstruction uncertainty with LOFAR, around 17 g/cm$^2$.\\}

We have probed the effects
of humidity on the lateral distribution of radio power by comparing two profiles
with high and low humidity. We performed this study for different frequency bands. 
In the LOFAR frequency band of 30--80~MHz the relative difference in power is small. 
For a higher frequency band of 50--350~MHz the same effects are comparatively
larger, up to 10$\%$. We also estimated the radiation
energy from the LDF profiles to see the effects of humidity on the reconstructed energy.
No significant difference was found for either frequency regime which indicates that humidity 
does not influence the estimated energy.
A Monte Carlo study on the reconstructed $X_{\rm max}$  
was also done for these frequency bands. No significant effect of humidity
is found on the reconstructed $X_{\rm max}$  for the low frequency band relevant for LOFAR; for the higher frequency band a mean difference on the order of 7 $\mathrm{g/cm^{2}}$ 
is observed. This could be important for the high precision $X_{\rm max}$  measurements for the cosmic ray detection with the SKA experiment \cite{ska}.\\

\textcolor{black}{In the process of implementing GDAS-based parameterized density and refractive index profile in CORSIKA/CoREAS, we have developed a tool, 
called \textquoteleft{gdastool}\textquoteright, which has been available for public use since the release of CORSIKA
version 7.6300, and is already being used by other experiments in the community around the globe.} \\

\textcolor{black}{In the previous LOFAR analysis the effects of refractive index  were included within the systematic uncertainties on the reconstructed $X_{\rm max}$. 
The improved atmospheric correction will lead to a reduced systematic uncertainty. An update on the  mass composition results 
is not within the scope of this study. It will be discussed in a future publication, 
which involves, along with atmospheric corrections, improved calibration of the radio antennas, energy scale, and new
$X_{\rm max}$ reconstruction techniques.}

\small
{\section{Acknowledgement}
The LOFAR cosmic ray key science project acknowledges funding from an Advanced Grant
of the European Research Council (FP/2007-2013)/ERC Grant Agreement no 227610. The project
has also received funding from the European Research Council (ERC) under the European Union's Horizon 2020 research and innovation program (grant agreement No 640130). We furthermore
acknowledge financial support from FOM, (FOM-project 12PR304). AN is supported by the DFG (Emmy-Noether grant NE 2031/2-1 ). LOFAR, the Low
Frequency Array designed and constructed by ASTRON, has facilities in several countries, that are
owned by various parties (each with their own funding sources), and that are collectively operated
by the International LOFAR Telescope foundation under a joint scientific policy. We sincerely thank
the CORSIKA developers for their assistance regarding the implementation of our work in CORSIKA modules.}
\small

\bibliographystyle{unsrt}
\bibliography{LOFARbib}

\end{document}